\documentstyle{amsppt}
\def\pt{\partial}
\def\kappa{\varkappa}
\vsize 14.5cm
\hsize 11cm
\TagsOnRight
\topmatter
\title On lump instability of Davey--Stewartson II equation
\endtitle
\author
Rustem R. Gadyl'shin and Oleg M. Kiselev
\endauthor
\affil
Institute of Mathematics, Ufa Science Center of Russian Academy of Sciences,
112, Chernyshevskii, 450000 Ufa, Russia
\endaffil
\email
grr\@imat.rb.ru
\endemail
\address\endaddress
\email
ok\@imat.rb.ru
\endemail
\abstract
We show that lumps (solitons) of the Davey--Stewartson II equation
fail under small perturbations of initial data.
\endabstract
\thanks
{Work partially supported by RFFI grant 97-01-00459}
\endthanks
\endtopmatter
\document
\head\S 1. Introduction
\endhead
\par
The Davey--Stewartson II (DS-II) equation \cite{1,2} is the well known
example
of the $2+1$ dimensional completely integrable equation (\cite{3,4}).
We study the lump (soliton) stability  for this equation. The DS-II
equation is
considered in the following form \cite{4}:
$$
\eqalign{
&
i\partial_t q+2(\partial^2_z+\partial^2_{\bar z})q+(g+\overline g)q=0,\cr
&
\partial_{\bar z}g=\partial_z| q|^2.}
$$
Here $z\in\Bbb C$, and the overbar represents complex
conjugation.
\par
It is known (\cite{4}) that there exists the correspondence
one-to-one between the lumps and poles (with respect to $k$) of
the matrix solution of the equation
$$
(I-G[q^0,k])\phi=E(kz),
\tag 1.1
$$
where $E(kz)=\hbox{\rm diag}(\exp(kz),\exp(\overset{\hrulefill}\to{kz}))$,
$I$ is the unit matrix, $q^0(z)$ is the initial data $q(z,0)$ for
the DS-II equation and
$$
\eqalign{
\
G[f,k]\phi\equiv
&
{i\over 4\pi}\int\int_{\Bbb C}d\zeta\wedge d\overline{\zeta}
\cr
&\times
\left(\matrix 0 &{f(\zeta)\exp(k(z-\zeta))\over z-\zeta}\cr
                {-\overline{f}(\zeta)\exp(\overline{k(z-\zeta)})\over
        \overline{z-\zeta}} & 0
\endmatrix\right)\phi(k,\zeta).
}
$$
Hereafter, the dependence of functions on the complex conjugate
va\-ri\-ab\-les is omitted.
\par
If the solution of (1.1) has a pole then the
solution of DS-II has the soliton structure.
We show that the pole disappears under small perturbation of the initial
data. One and the same the soliton structure is instable with respect to
perturbations. Namely in this sence we mean instability of the lump.
\par
\head
\S 2. Reduction of the problem
\endhead
The existence of the nontrivial solution for the boundary value pro\-blem
$$
\aligned
&
(I-G[q^0,k_0])\Phi=0,\\
&
E(-k_0z)\Phi(z)\to 0,
\qquad |z|\to\infty
\endaligned\tag2.1
$$
is the nessecary condition for existence
of the pole at $k=k_0$ for the solution of (1.1) (see, for instance,
\cite{4}). In turn, one can see,
that the existence of the nontrivial solutions of (2.1) is equivalent to
the existence of the nontrivial solution (vanishing at infinity)
of the equation
$$
(I-\Cal G[q^0,k_0])\chi=0,\tag2.2
$$
where
$$
\eqalign{
{\Cal {\Cal G}}[f,k]\chi=
&
{i\over 4\pi}\int\int_{\Bbb C}
d\zeta\wedge d\overline{\zeta}
\cr
&\times
\left(\matrix 0 &{f(\zeta)\exp(\overline{k\zeta}-k\zeta)\over z-\zeta}\cr
                {-\overline{f}(\zeta)\exp(k\zeta-\overline{k\zeta})\over
        \overline{z-\zeta}} & 0
\endmatrix\right)\chi(k,\zeta).
}
$$
\par
One can show, that if $q^0$
is a continuous function and $|q^0|\sim |z|^{-2}$ as $|z|\to \infty$,
then ${\Cal G}[q^0,k]$ is the compact operator
in the space $\Cal C(\Bbb C)$ of continuous vector-functions with the norm
$$
|| u||=\sup\limits_{z\in\Bbb C}(|u_1|+|u_2|),
$$
where $u_j$ are the components of $u$  (see also, \cite{5}).
So, the existence of the lumps is reduced to the
existence of zeros for eigenvalues of the equation
$$
(I-{\Cal G}[q^0,k])A=\lambda(k)A\tag2.3
$$
\par
We show that zero of the eigenvalue is instable with respect to
a small perturbation of the potential $q^0$. As above mentioned this
result implies the instability of the lumps.
\head
\S 3. Uniform asymptotics of the eigenvalue
\endhead
\par
Assume that the solution of
(1.1) has the first order
pole at $k=k_0$ in the unperturbed case $q^0=q_0$ and denote

$$
\sigma = \left(\matrix 0 & 1\cr
                                -1 & 0
                        \endmatrix\right).
$$
It is known \cite{4}, that if
$A^{(1)}$
is the nontrivial solution of (2.2) (for $q^0=q_0(z)$), then
$A^{(2)}=\sigma \overline{A^{(1)}}\sigma^{-1}$
is the second nontrivial solution of (2.2).
Obviously, $A^{(i)}$ are
the eigenfunction of (2.3) for $k=k_0$, $q^0=q_0$ and $\lambda(k_0)=0$.
To be more concrete, we assume  that the multiplicity of the null
eigenvalue, for $k_0=0$, is equal to two.
\par
Now, we consider the perturbed case
$$
q^0(z)=q_0(z)+\varepsilon q_1(z),\tag3.1
$$
where $q_1$ is a smooth function with the finite support and
$\varepsilon$ is a small real valued parameter.
The leading terms of asymptotics
for the eigenvalue and eigenfunction of (2.3) in
the perturbed case are constructed in the following form:
$$
\allowdisplaybreaks
\align
\tilde\lambda(\varepsilon,k)=
&
\varepsilon \overset{1}\to{\lambda}(\varkappa),
\tag 3.2\\
\tilde A(z,k,\varepsilon)=
&
\overset{0}\to{A}(z,\varkappa)+\varepsilon\overset{1}
\to{A}(z,\varkappa),
\tag 3.3
\endalign
$$
where $\kappa=(k-k_0)\varepsilon^{-1}$.
Substituting (3.1)-(3.3) into (2.3), we obtain the following
equations:
$$
\allowdisplaybreaks
\align
&
(I-{\Cal G}[q_0,k_0])\overset{0}\to{A}=0,\tag3.4\\
&
(I-{\Cal G}[q_0,k_0])\overset{1}\to{A}=\overset 1\to F,\tag3.5
\endalign
$$
where
$$
\overset 1\to F={\Cal G}[q_1,k_0]\overset{0}\to{A}+
\kappa\pt_k {\Cal G}[q_0,k_0]\overset{0}\to{A}+
\bar\kappa\pt_{\bar k}{\Cal G}[q_0,k_0]\overset{0}\to{A}+
\overset{1}\to{\lambda}(\kappa)\overset{0}\to{A}.\tag3.6
$$
Obviously, that
$$
\overset{0}\to{A}(z,\varkappa)=\alpha(\varkappa)A^{(1)}(z)+\beta
(\varkappa)A^{(2)}(z)
\tag3.7
$$
satisfies (3.4) for any $\alpha$ and $\beta$.
\par
Denote
$$
(u,v)_f=
\int\int_{\Bbb C} dz\wedge d\overline{z}\,\left(
\overline{f}\exp(kz-\overline{kz})\,u_1\overline{v_1}+
        f\exp(\overline{kz}-kz)\,u_2\overline{v_2}\right),
$$
where $u_i,\,v_i$ are the components of the vectors $u,\,v$ respectively.
Let $B^{(1)}$ be the eigenfunction of
the formal adjoint equation for
$(I+\Cal G[q_0,k_0])$  with respect to $(\bullet,\bullet)_{q_0}$.
One can see, that the function
$$
B^{(2)}=\sigma \overline{B^{(1)}}\sigma^{-1}
$$
is the second eigenfunction of the adjoint equation.
The Fredholm alternative for the solvabililty of (3.5)
reads as following:
$$
(\overset{i}\to{F},B^{(j)})_{q_0} = 0,\qquad j=1,2.\tag3.8
$$
We normalize $A^{(1)}$  and  $B^{(1)}$ by conditions
$$
(A^{(1)},e_1)_{q_0}=0,\quad (A^{(1)},e_2)_{q_0}=4i\pi,
$$
$$
(e_1,B^{(1)})_{q_0}=0,\quad (e_2,B^{(1)})_{q_0}=4i\pi,
$$
where $e_i,\,i=1,2$ is the corresponding column of the unit matrix.
\par
The  construction of the asymptotics is standard.
Putting (3.6) in (3.8),  we  get the following equations
$$
\aligned
&
\overset{1}\to{\lambda}\alpha\Omega_1+\overset{1}\to{\lambda}\beta\Omega_2
+\alpha Q_1+\beta Q_2+4i\pi\beta\bar\kappa=0,\\
&
\overset{1}\to{\lambda}\alpha\bar\Omega_2-
\overset{1}\to{\lambda}\beta\bar\Omega_1+
\alpha \bar Q_2-\beta \bar Q_1-4i\pi\alpha\kappa=0,
\endaligned\tag3.9
$$
where
$$
\Omega_j=-4i\pi(A^{(j)},B^{(1)})_{q_0},\qquad
Q_j=(A^{(1)},B^{(j)})_{q_1},\qquad j=1,2.
$$
The system (3.9) implies that
$$
\eqalign{
&
\left(|\Omega_1|^2+|\Omega_2|^2\right)\overset{1}\to{\lambda}^2+
2\hbox{\rm Re}
\left(\overline{\Omega_1}Q_1+\Omega_2\left(\overline{Q_2}-4i\pi
\bar\kappa\right)\right)\overset{1}\to{\lambda}+
\cr
&+
|4i\pi\bar\kappa+Q_2|^2+|Q_1|^2=0,
}
\tag3.10
$$
The equation (3.10) has two solutions $\overset{1}\to{\lambda}_i,\,i=1,2,$
(taking in account their multiplicity) such that
$$
\allowdisplaybreaks
\align
&
\overset{1}\to{\lambda}_1(\varkappa)
\overset{1}\to{\lambda}_2(\varkappa)=
{|4i\pi\bar\kappa+Q_2|^2+|Q_1|^2\over|\Omega_1|^2+|\Omega_2|^2},
\tag3.11\\
&
\overset{1}\to{\lambda}_i(\varkappa)=O(1+|\varkappa|).
\tag3.12
\endalign
$$
Hereafter, the estimates cointainig $\varkappa$ are understood as uniform
with respect to $\varkappa\in\Bbb C$.
\par
\remark{Remark 3.1} The following constructions are
independent of the index $i$ in the formulae
$\overset{1}\to{\lambda}_i$. Therefore we omit it in the
construction of the eigenvalues and eigenfunctions up to remark 3.2 below.
\endremark
\par
Let us normalize (3.7) by condition:
$$
\alpha+|\beta|=1,\qquad \alpha\ge0\tag3.13
$$
Due to (3.13) and any equation from (3.9),
we get $\alpha$, $\beta$ and  obtain the
solution of (3.5) such that
$$
||\overset{1}\to A(z,\varkappa)||=O(1+|\varkappa|).\tag3.14
$$\par
\par
Formalae (3.4)-(3.7) and (3.12)-(3.14) imply that there exist
$\tilde\lambda$ and $\tilde A$
having form (3.2) and (3.3), respectively, such that
$$
T(\varepsilon,k)
\tilde A=\tilde F,\tag3.15
$$
where $T(\varepsilon,k)=I-{\Cal G}[q,k]-\tilde\lambda(\varepsilon,k)$,
$$
\align
&
0<C_1<||\tilde A||<C_2,\tag3.16\\
&
||\tilde F||=o(|\varepsilon|+
|k-k_0|).\tag3.17
\endalign
$$
Hereafter $C_i$ are constants.
Obviously, for the eigenvalues $\mu(\varepsilon,k)$ of
$T(\varepsilon,k)$ and
$\lambda(\varepsilon,k)$ of $I-{\Cal G}[q,k]$ there is the equality
$$
\lambda(\varepsilon,k)=\tilde\lambda(\varepsilon,k)+\mu(\varepsilon,k).
\tag3.18
$$
\par
Let $\gamma$ be the circle around the origin of enough small fixed
radius. Denote
$$
\allowdisplaybreaks
\align
P(\varepsilon,k)
=
&
-{1\over2\pi i}\int\limits_\gamma
\left(T(\varepsilon,k)-\zeta\right)^{-1}d\zeta,\\
U(\varepsilon,k)
=
&
\left(1-\left(P(\varepsilon,k)-P(0,k_0)\right)^2\right)^{-1/2}
\left(1-P(\varepsilon,k)-P(0,k_0)\right).
\endalign
$$
It is known \cite{6}, that the projector
$P(\varepsilon,k)$ and transform function
$U(\varepsilon,k)$ reduce the eignevalue problem (local) for
$T(\varepsilon,k)$ to the the eignevalue problem for the two dimensional
operator
$$
\hat T(\varepsilon,k)=
U^{-1}(\varepsilon,k)T(\varepsilon,k)U(\varepsilon,k)P(0,k_0)
$$ defined
in $P(0,k_0)\Cal C(\Bbb C)$. Namely, both eigenvalues of $\hat T$
and both vanishing
eigenvalues of $T$ (as $\varepsilon\to0$ and $k\to k_0$) coincide.
By definition of $T$ the operator $\hat T$ has form
$$
\hat T(\varepsilon,k)=\varepsilon \hat T_1(\varepsilon,k)+
(k-k_0)\hat T_2(\varepsilon,k),\qquad||T_i||\le C,\tag3.19
$$
and due to (3.15)-(3.17) the function
$\hat A(\varepsilon,k)=U^{-1}(\varepsilon,k)\tilde A(\varepsilon,k)$
satisfies the equation
$$
\hat T(\varepsilon,k)
\hat A=o(|\varepsilon|+|k-k_0|)\tag3.20
$$
and the following estimate holds
$$
0<C_3<||\hat A||< C_4.\tag3.21
$$
\par
Suppose, that
$$
k-k_0=O(\varepsilon).\tag3.22
$$
and denote $B(\varepsilon,k)=\varepsilon^{-1}\hat T(\varepsilon,k)$.
Let $\nu(\varepsilon,k)$ is the eigenvalue of $B(\varepsilon,k)$.
Obviously, equality (3.18) implies that
$$
\lambda(\varepsilon,k)=\tilde\lambda(\varepsilon,k)+\varepsilon
\nu(\varepsilon,k).\tag3.23
$$
On the other hand, due to (3.19), (3.20) and (3.22)  we have that
$$
B(\varepsilon,k)\tilde A=o(1),\qquad ||B||\le C_5.\tag3.24
$$
In turn, formulae (3.21) and (3.24) imply that there exists the vanihing
$\nu(\varepsilon)$ and, hence (see, (3.23)) there is the eigenvalue
$$
\lambda(\varepsilon,k)=\tilde\lambda(\varepsilon,k)+o(\varepsilon).\tag3.25
$$
\par
In the same way, one can obtain that if
$$
\varepsilon=O(k-k_0)\tag3.26
$$
then
$$
\lambda(\varepsilon,k)
=\tilde\lambda(\varepsilon,k)+o((k-k_0)).\tag3.27
$$
\par
Formulae (3.22), (3.25)--(3.27) imply that
$$
\lambda(\varepsilon,k)
=\tilde\lambda(\varepsilon,k)+o(|k-k_0|+|\varepsilon|),\qquad
k-k_0,\,\varepsilon\to0.\tag3.28
$$
\par
\remark{Remark 3.2} Now, recall that, in fact,  we
constructed two series a\-symp\-to\-tics cor\-re\-s\-pon\-d\-ing the index
$i=1,2$ (see,
remark 3.1). So, below we add this index for the constructed quantities.
\endremark
\par
Denote by $\lambda_i$, $i=1,2$, the eigenvalues of
$(I-{\Cal G}[q^0,k])$ vanishing as $k\to k_0$ and
$\varepsilon\to0$. Due to (3.2), (3.11), (3.12) and (3.28)
we have that
$$
\lambda_1\lambda_2=
{|4i\pi(\overline{k-k_0})+\varepsilon Q_2|^2+\varepsilon^2
|Q_1|^2\over|\Omega_1|^2+|\Omega_2|^2}+o(\varepsilon^2+
|\varepsilon(k-k_0)|+|k-k_0|^2).
$$
One can see that, if $Q_1\not=0$, then
$\lambda_1\lambda_2\not=0$ for $\varepsilon>0$
and any $k$ close to $k_0$.
\par\remark{Remark 3.3} So, the structure of the
scattering data (in the terms of the inverse scattering transform \cite{4})
is instable under small perturbations of initial data.
This result justifies the formal asymptotics of the continuous part for
the scattering data constructed in \cite{5}.
\endremark


\medskip
\Refs
\ref\no 1
\by Davey A., Stewartson K.
\paper On the three-dimensional packets of surfase waves
\jour Proc. R. Soc. London. Ser.A.
\vol 338
\yr 1974
\pages 101--110
\endref
\ref\no 2
\by Djordjevic V.D., Redekopp L.G.
\paper On two-dimensional packets of capillary- gravity waves
\jour J.Fluid Mech.
\vol 79
\yr 1977
\pages 703--714
\endref
\ref\no 3
\by Fokas A.S., Ablowitz M.J.
\paper On the inverse scattering transform of mul\-ti\-di\-men\-sional nonlinear
equations related to first-order system in the plane.
\jour J. Math. Phys.
\yr 1984
\vol 25
\pages 2494--2505
\endref
\ref\no 4
\by Arkadiev V.A., Pogrebkov A.K., Polivanov M.C.
\paper Inverse scattering transform method and soliton solutions for
Davey-Stewartson II equation.
\jour Physica D.
\yr 1989
\vol 36
\pages 189-197
\endref
\ref\no 5
\by Gadyl'shin R.R., Kiselev O.M.
\paper On nonsoliton structure of scattering data under prturbation
of two-dimentional soliton for Davey -- Stewartson equation II
\jour Teor. i Mat. Fiz.
\yr 1996
\vol 106
\pages 200-208
\endref
\ref \no 6
\by Kato T.
\book Perturbation Theory for Linear Operators
\yr 1966
\publaddr
\publ Springer--Verlag
\endref
\endRefs
\enddocument
\end